\documentstyle[manuscript,eqsecnum,aps]{revtex}

\begin{document}


\preprint{IMPERIAL/TP/95-96/51}

\title{Multiple-Scale Analysis of Quantum Systems}

\author{Carl M. Bender\footnote{Permanent address: Department of Physics,
Washington University, St. Louis, MO 63130, USA}
and Lu\'{\i}s M. A. Bettencourt}
\address{Blackett Laboratory, Imperial College, London SW7 2BZ, United Kingdom}

\date{July 9, 1996}

\maketitle

\begin{abstract}
Conventional weak-coupling Rayleigh-Schr\"odinger perturbation theory suffers
from problems that arise from resonant coupling of successive orders in the
perturbation series. Multiple-scale analysis, a powerful and sophisticated
perturbative method that quantitatively analyzes characteristic physical
behaviors occurring on various length or time scales, avoids such problems by
implicitly performing an infinite resummation of the conventional perturbation
series. Multiple-scale perturbation theory provides a good description of the
classical anharmonic oscillator. Here, it is extended to study (1) the
Heisenberg operator equations of motion and (2) the Schr\"odinger equation for
the quantum anharmonic oscillator. In the former case, it leads to a system of
coupled operator differential equations, which is solved exactly. The solution
provides an operator mass renormalization of the theory. In the latter case,
multiple-scale analysis elucidates the connection between weak-coupling
perturbative and semiclassical nonperturbative aspects of the wave function.
\end{abstract}

\pacs{PACS number(s): 11.15.Bt, 02.30.Mv, 11.15.Tk}

\section{INTRODUCTION}
\label{s1}
Multiple-scale perturbation theory (MSPT) is a powerful and sophisticated
perturbative technique for solving physical problems having a small parameter
$\epsilon$ \cite{BO,Nay,MS}. This perturbation method is generally useful for
both linear and nonlinear problems. In fact, it is so general that other
well-known perturbative methods, such as WKB theory and boundary-layer theory,
which are useful in more limited contexts, may be viewed as special cases of
MSPT \cite{BO}.

The key idea underlying MSPT is that dynamical
systems tend to exhibit distinct characteristic physical behaviors at different
length or time scales. If a conventional perturbation series is used to solve a
problem, then there is often a resonant coupling between successive orders of
perturbation theory. This coupling gives rise to {\em secular terms} in the
perturbation series (terms that grow rapidly as functions of the length or time
variable). These secular terms conflict with physical requirements that the
solution be finite. MSPT {\em reorganizes} the perturbation series to eliminate
secular growth, and in doing so it gives a quantitative description of the
characteristic behaviors that occur at many scales. In the past MSPT has been
used to solve {\em classical} differential equations such as the equation of
motion for the classical anharmonic oscillator. Indeed, the classical anharmonic
oscillator is often used to illustrate and explain the method of MSPT.

In this paper we generalize the ideas of MSPT to the {\em quantum} anharmonic
oscillator \cite{BB}. The quantum anharmonic oscillator is an excellent
laboratory for the study of a variety of perturbative methods. It has been used
to study the origin of the divergence of conventional weak-coupling
Rayleigh-Schr\"odinger perturbation theory \cite{BW1}, Pad\'e and Borel
summation of perturbation series \cite{Simon}, large-order behavior of
perturbation theory \cite{BW2}, delta expansions \cite{DELTA}, dimensional
expansions \cite{DIML}, and strong-coupling expansions \cite{STRONG}. Here, we
use MSPT to study two aspects of the quantum anharmonic oscillator, the
Heisenberg operator equations of motion in Secs.~\ref{s2} and \ref{s3} and the
Schr\"odinger equation in Secs.~\ref{s4} and \ref{s5}. We illustrate the methods
of MSPT in Sec.~\ref{s2} by applying it to the nonlinear dynamical equation of
motion for the classical anharmonic oscillator (Duffing's equation). There, we
obtain the first-order frequency shift. Then, in Sec.~\ref{s3} we extend the
methods of MSPT to the nonlinear Heisenberg equation for the
quantum anharmonic oscillator (the quantum version of Duffing's equation). To
complete the analysis it is necessary to solve a nonlinear system of coupled
operator differential equations. We find the exact closed-form solution to this
system. From this solution, we obtain the quantum operator analog of the
frequency shift; namely, an operator mass renormalization that expresses the
first-order shift of all energy levels.

In the next two sections we study the Schr\"odinger equation for the quantum
anharmonic oscillator. Specifically, we examine the asymptotic behavior of the
wave function $\psi(x)$ for large $x$. We consider the problem of reconciling
the different results that one obtains from conventional Rayleigh-Schr\"odinger
perturbation theory (a formal Taylor series in powers of a small parameter
$\epsilon$) and WKB theory (a nonperturbative probe of the anharmonic oscillator
that is valid regardless of the size of $\epsilon$). To any finite order in
conventional perturbation theory, $\psi(x)$ behaves like a Gaussian for large
$x$; however, WKB theory predicts that as $x\to\infty$ the wave function decays
to zero like the exponential of a cubic. In Sec.~\ref{s4} we resolve this
discrepancy at two different length scales by an infinite sequence of
reorderings and resummations of the conventional weak-coupling perturbation
series. In Sec.~\ref{s5} we explain the origin of the disparity by performing a
direct multiple-scale analysis of the Schr\"odinger equation for the quantum
anharmonic oscillator.

The approach used in this paper for the anharmonic oscillator wave function has
been applied in perturbative quantum field theory to sum leading-logarithm
divergences \cite{CW} and leading infrared divergences \cite{DJP}. It is our
hope that in the future the direct nonperturbative multivariate approach of MSPT
will provide a framework to simplify such schemes.

\section{MULTIPLE-SCALE PERTURBATION THEORY APPLIED TO THE CLASSICAL ANHARMONIC
OSCILLATOR}
\label{s2}
In this section we explain MSPT by using it to treat the classical anharmonic
oscillator, a dynamical system satisfying the nonlinear differential equation
\begin{equation}
{d^2\over d t^2}y+y+4\epsilon y^3=0\quad (\epsilon>0),
\label{e2.1}
\end{equation}
which is known as Duffing's equation. The positivity of $\epsilon$ ensures that
there are no unbounded runaway modes. We impose the initial conditions
\begin{equation}
y(0)=1\quad{\rm and}\quad y^{\prime}(0)=0.
\label{e2.2}
\end{equation}

The harmonic oscillator ($\epsilon=0$) has only one time scale, namely, the
period of oscillation. However, the nonlinear term in Eq.~(\ref{e2.1})
introduces many time scales into the problem. For example, when $\epsilon\neq
0$, one can observe on a long-time scale [$t={\rm O}(\epsilon^{-1})$] a
frequency shift of order $\epsilon$. One can study the classical anharmonic
oscillator on the short-time scale $t$ and also on many long-time scales
$\tau=\epsilon t$, $\tau_1=\epsilon^2 t$, $\tau_2=\epsilon^3 t$, and so on.

Let us first examine what happens if we attempt to solve Duffing's equation
using a conventional perturbation series in powers of the parameter $\epsilon$,
\begin{equation}
y(t) = \sum_{n=0}^{\infty} \epsilon^n y_n(t),
\label{e2.3}
\end{equation}
in which the initial conditions in Eq.~(\ref{e2.2}) are contained as
\begin{eqnarray}
y_0(0) &=& 1\quad{\rm and}\quad y^\prime_0(0)=0,\nonumber\\
y_n(0) &=& y^\prime_n(0)=0 \quad (n\geq 1).
\label{e2.4}
\end{eqnarray}
Substitute Eq.~(\ref{e2.3}) into Eq.~(\ref{e2.1}). To leading order (zeroth
order in powers of $\epsilon$), we have
\begin{equation}
{d^2\over d t^2}y_0+y_0=0
\label{e2.5}
\end {equation}
and to first order in powers of $\epsilon$ we have
\begin{equation}
{d^2\over d t^2}y_1+y_1=-4y_0^3.
\label{e2.6}
\end {equation}

The solution to Eq.~(\ref{e2.5}) satisfying the initial conditions in
Eq.~(\ref{e2.4}) is
\begin{equation}
y_0(t)=\cos t.
\label{e2.7}
\end{equation}
When we introduce this solution into Eq.~(\ref{e2.6}), we obtain
\begin{equation}
{d^2\over d t^2}y_1+y_1=-\cos(3 t)-3\cos t.
\label{e2.8}
\end {equation}
Equation (\ref{e2.8}) represents a forced harmonic oscillator whose driving term
has frequencies 3 and 1. When a harmonic oscillator is driven at its natural
frequency, which in this case is 1, we have the phenomenon of resonance. As a
result, the solution
\begin{equation}
y_1(t) = {1 \over 8}\cos(3 t) -{1 \over 8}\cos t - {3 \over 2} t\sin t
\label{e2.9}
\end{equation}
contains a secular term that grows linearly with increasing time $t$.
Equation~(\ref{e2.9}) cannot be valid for long times because the exact solution
to Duffing's equation remains bounded for all $t$ \cite{BO}. Hence, the
conventional perturbation expansion is sensible only for short times $t<<
\epsilon^{-1}$. How then does the conventional perturbation series determine the
behavior of $y(t)$ for long times, say of order $\epsilon^{-1}$? 

One way to answer this question is to identify the structure of the most secular
(highest power in $t$) term {\em to all orders} in perturbation theory. One can
easily verify \cite{BO} that for all $n$ the most secular term in $y_n(t)$ has
the form
\begin{equation}
{1\over 2}{t^n\over n!}\left[\left({3i\over 2}\right)^ne^{it}+\left(-{3i\over 2}
\right)^n e^{-it}\right].
\label{e2.10}
\end{equation}
Since the expression in Eq.~(\ref{e2.10}) is multiplied by $\epsilon^n$, if we
make the approximation of retaining only the most secular term in every order,
then we obtain a series in powers of the long-time variable $\tau=\epsilon t$.
Evidently, we can sum the most secular terms in this series to all orders in
$\epsilon$, and since the result is a cosine function, it remains bounded for
all times $t$:
\begin{equation}
{1\over 2}\sum_{n=0}^{\infty}{\epsilon^n t^n\over n!}\left[\left({3i\over 2}
\right)^ne^{it}+\left(-{3i\over 2}\right)^ne^{-it}\right]
=\cos\left[\left(1+{3\over2}\epsilon\right)t\right].
\label{e2.12}
\end{equation}
We interpret this result to mean that on the long-time scale $\tau$ there is a 
{\em frequency shift} in the oscillator of order ${3\over 2}\epsilon$. Of
course, this result is not exact; there are less secular terms to all orders in
the perturbation expansion, and these terms give rise to frequency shifts of
order $\epsilon^2$, $\epsilon^3$, and so on.

We will now show how MSPT directly reproduces the
result in Eq.~(\ref{e2.12}). To avoid the complicated procedure of summing the
conventional perturbation series to all orders in powers of $\epsilon$, MSPT
uses a sophisticated perturbative approach that prevents secular terms from
appearing in the perturbation expansion. Multiple-scale analysis assumes {\em a
priori} the existence of many time scales ($t$, $\tau$, $\tau_1$, $\tau_2$,
$...$) in the problem, which can be temporarily treated as {\em independent}
variables. Here, we illustrate by performing just a first-order calculation. We
use only the two variables $t$ and $\tau=\epsilon t$ and seek a perturbative
solution to Eq.~(\ref{e2.1}) of the form
\begin{equation}
y(t)=Y_0(t,\tau)+\epsilon Y_1(t,\tau)+{\rm O}(\epsilon^2).
\label{e2.13}
\end{equation}

Using the chain rule and the identity ${\partial\tau\over\partial t}=\epsilon$,
we convert Eq.~(\ref{e2.1}) to a sequence of {\em partial} differential
equations for the dependent variables $Y_0$, $Y_1$, $...~$. The first two
equations read
\begin{eqnarray}
{\partial^2\over \partial t^2}Y_0+Y_0 &=& 0,
\label{e2.14} \\
{\partial^2\over \partial t^2}Y_1+Y_1 &=& -4Y_0^3-2 
{\partial^2\over\partial t\partial\tau}Y_0.
\label{e2.15}
\end {eqnarray}
The general solution to Eq.~(\ref{e2.14}) is 
\begin{equation}
Y_0 (t,\tau)=A(\tau)\cos t +B(\tau)\sin t,
\label{e2.16}
\end{equation}
where $A(\tau)$ and $B(\tau)$ are as yet unknown functions of $\tau$. We
determine these functions by imposing the condition that no secular terms appear
in $Y_1$. That is, we use the functional freedom in the choice of $A(\tau)$ and
$B(\tau)$ to eliminate the resonant coupling between zeroth and first order in
perturbation theory. We substitute $Y_0(t,\tau)$ in Eq.~(\ref{e2.16}) into the
right side of Eq.~(\ref{e2.15}) and expand the resulting expression using the
trigonometric identities
\begin{eqnarray}
\cos^3 t &=& {1\over 4}\cos(3t) +{3\over 4}\cos t,\nonumber\\
\sin^3 t &=& -{1\over 4}\sin(3t) +{3\over 4}\sin t,\nonumber\\
\cos^2 t\,\sin t &=& {1\over 4}\sin(3t) +{1\over 4}\sin t,\nonumber\\
\sin^2 t\,\cos t &=& -{1\over 4}\cos(3t) +{1\over 4}\cos t.
\label{id}
\end{eqnarray}
To eliminate secularity we require that the coefficient of $\cos t$ and $\sin t$
vanish. This gives the pair of equations
\begin{equation}
2{dB\over d\tau}=-3A^3-3AB^2\quad{\rm and}\quad 2{dA\over d\tau}=3B^3+3A^2B.
\label{e2.17}
\end{equation}

To solve this system we multiply the first equation by $B(\tau)$ and the second
equation by $A(\tau)$. Adding the resulting equations gives
\begin{equation}
{d\over d\tau}C(\tau)=0,
\label{e2.18}
\end{equation}
where
\begin{equation}
C(\tau)={1\over 2}\left[ A(\tau)\right]^2+{1\over 2}\left[ B(\tau)\right]^2.
\label{e2.19}
\end{equation}
Hence $C(\tau)$ is independent of  $\tau$ and we may take $C(\tau)=C(0)$.

Substituting this result back into Eq.~(\ref{e2.17}) gives the elementary linear
system
\begin{eqnarray}
{d\over d\tau}B=-3C(0) A\quad {\rm and}\quad{d\over d\tau}A=3C(0)B.
\label{e2.20}
\end{eqnarray}
When we solve this system and then impose the initial conditions, we obtain
$C(0)={1\over 2}$ and
\begin{equation}
Y_0(t,\tau)=\cos\left[(1+{3\over 2}\epsilon)t\right],
\label{e2.21}
\end{equation}
where we have used $\tau=\epsilon t$.
We have thus reproduced the approximate solution in Eq.(\ref{e2.12}) that is
valid for long times. Note that while conventional perturbation theory is only
valid for $t<<\epsilon^{-1}$, the multiple-scale solution is valid for times
satisfying $t<<\epsilon^{-2}$. This multivariate approach can, in principle, be
performed to $n$th order in powers of $\epsilon$ for any $n$.

\section{MULTIPLE-SCALE PERTURBATION THEORY APPLIED TO THE QUANTUM ANHARMONIC
OSCILLATOR}
\label{s3}

In this section we extend the multiple-scale approach described in Sec.~\ref{s2}
to the quantum anharmonic oscillator. This is a nontrivial generalization of the
usual multiple-scale techniques because it requires that we solve {\em operator}
differential equations \cite{OPEQNS}.

The quantum anharmonic oscillator is defined by the Hamiltonian \cite{MASS}
\begin{equation}
H(p,q)={1\over 2}p^2+{1\over 2}q^2+\epsilon q^4 \quad (\epsilon >0),
\label{e3.1}
\end{equation}
where $p$ and $q$ are operators satisfying the canonical equal-time 
commutation relation
\begin{equation}
[q(t),p(t)]=i\hbar .
\label{e3.2}
\end{equation} 
The positivity of $\epsilon$ ensures that $H(p,q)$ is bounded below.

The Heisenberg operator equations of motion are
\begin{eqnarray}
{d\over dt}q &=& {1\over i\hbar}[q,H(p,q)] =p,
\label{e3.3}\\
{d\over dt}p &=& {1\over i\hbar}[p,H(p,q)]=-q(t)-4\epsilon[q(t)]^3.
\label{e3.4}
\end{eqnarray}
These equations combine to give
\begin{eqnarray}
{d^2\over dt^2}q(t)+q(t)+4\epsilon [q(t)]^3=0,
\label{e3.5} 
\end{eqnarray}
which is the quantum analog of Duffing's classical differential equation
(\ref{e2.1}). Since $p(t)$ and $q(t)$ are operators, we cannot impose numerical
initial conditions like those in Eq.~(\ref{e2.2}); rather, we enforce a general
operator initial condition at $t=0$:
\begin{equation}
q(0) = q_0 \quad {\rm and}\quad p(0) = p_0.
\label{e3.6}
\end{equation}  
Here, $p_0$ and $q_0$ are fundamental time-independent operators obeying the
Heisenberg algebra
\begin{equation}
[q_0, p_0]=i\hbar . 
\label{e3.7}
\end{equation}  

Rather than seeking a solution to Eq.~(\ref{e3.5}) as a conventional
perturbation series in powers of $\epsilon$, we perform a multiple-scale
analysis. We assume that $q(t)$ exhibits characteristic behavior on the
short-time scale $t$ and on the long-time scale $\tau=\epsilon t$ and we write
\begin{equation}
q(t)=Q(t,\tau)=Q_0(t,\tau)+\epsilon Q_1(t,\tau)+{\rm O}(\epsilon^2).
\label{e3.8}
\end{equation}
This equation is analogous to Eq.~(\ref{e2.13}) but here $Q_0$ and $Q_1$ are
{\em operator-valued} functions.

There is an associated expression for the momentum operator $p(t)$:
\begin{eqnarray}
p(t)= P_0(t,\tau)+\epsilon P_1(t,\tau)+{\rm O}(\epsilon^2).
\label{e3.9}
\end{eqnarray}
Furthermore, since the momentum operator is the time derivative of the position
operator [see Eq.~(\ref{e3.3})], the chain rule gives
\begin{eqnarray}
p(t)= {\partial\over\partial t}Q_0+\epsilon \left({\partial\over\partial\tau}Q_0
+{\partial\over\partial t}Q_1\right)+{\rm O}(\epsilon^2).
\label{e3.10}
\end{eqnarray}

We substitute the expression for $q(t)$ in Eq.~(\ref{e3.8}) into
Eq.~(\ref{e3.5}), collect the coefficients of $\epsilon^0$ and $\epsilon^1$,
and obtain operator differential equations analogous to Eqs.~(\ref{e2.14}) and
(\ref{e2.15}):
\begin{eqnarray}
{\partial^2\over\partial t^2}Q_0+Q_0 &=& 0,\label{e3.11}\\
{\partial^2\over\partial t^2}Q_1+Q_1 &=& -4Q_0^3-2{\partial^2\over
\partial t\partial\tau}Q_0.
\label{e3.12}
\end{eqnarray}

Because Eq.~(\ref{e3.11}) is linear it is easy to find its general solution:
\begin{eqnarray}
Q_0(t,\tau) &=& {\cal A}(\tau)\cos t +{\cal B}(\tau)\sin t;
\label{e3.13}
\end{eqnarray} 
from this result we obtain
\begin{eqnarray}
P_0(t,\tau) &=& {\cal B}(\tau)\cos t -{\cal A}(\tau)\sin t.
\label{e3.14}
\end{eqnarray} 

It is now necessary to find the coefficient functions ${\cal A}(\tau)$ and
${\cal B}(\tau)$, which are operators. The canonical commutation relations in
Eq.~(\ref{e3.2}) imply that these operators must satisfy
\begin{equation}
[{\cal A}(\tau),{\cal B}(\tau)]=i\hbar .
\label{e3.15}
\end{equation}
Also, the initial conditions in Eq.~(\ref{e3.6}) give
\begin{equation}
{\cal A}(0) = q_0 \quad {\rm and}\quad {\cal B}(0) = p_0.
\label{e3.16}
\end{equation}

To determine the $\tau$ dependence of the functions ${\cal A}(\tau)$ and
${\cal B}(\tau)$, we must eliminate secular behavior. To do so we examine the
right side of Eq.~(\ref{e3.12}), which is 
\begin{equation}
-4\left[{\cal A}(\tau)\cos t +{\cal B}(\tau)\sin t\right]^3-2\left[-{d\over d
\tau}{\cal A}(\tau)\sin t +{d\over d\tau}{\cal B}(\tau)\cos t\right].
\label{e3.17}
\end{equation}
Next, we expand the cubic term, taking care to preserve the order of operator
multiplication, and we use the trigonometric identities in Eq.~(\ref{id}). We
eliminate secularity by setting the coefficients of $\cos t$ and $\sin t$ to
zero and obtain
\begin{eqnarray}
2{d{\cal B}\over d\tau} = -3{\cal A}^3
-{\cal B}{\cal A}{\cal B}-{\cal B}{\cal B}{\cal A}-{\cal A}{\cal B}{\cal B}
\quad{\rm and}\quad 2{d{\cal A}\over d\tau} = 3{\cal B}^3
+{\cal A}{\cal B}{\cal A}+{\cal A}{\cal A}{\cal B}+{\cal B}{\cal A}{\cal A}.
\label{e3.18}
\end{eqnarray}
This system of operator-valued differential equations is the quantum analog of
Eq.~(\ref{e2.17}).

To solve the system (\ref{e3.18}) we begin by multiplying the first equation on
the left and on the right by ${\cal B}(\tau)$ and the second equation on the
left and on the right by ${\cal A}(\tau)$. Adding the resulting four equations
and simplifying, we get
\begin{equation}
{d\over d\tau}{\cal H}=0,
\label{e3.19}
\end{equation}
where
\begin{equation}
{\cal H}\equiv {1\over 2}{\cal A}^2+{1\over 2}{\cal B}^2.
\label{e3.20}
\end{equation}
Equation (\ref{e3.19}) is the quantum analog of ${dC\over d\tau}=0$ in
Eq.~(\ref{e2.18}). By construction, the operator ${\cal H}$ is independent of
the short-time variable $t$. However, Eq.~(\ref{e3.19}) shows that ${\cal H}$ is
also independent of the long-time variable $\tau$. Therefore, Eq.~(\ref{e3.16})
allows us to express ${\cal H}$ in terms of the fundamental operators $p_0$ and
$q_0$:
\begin{equation}
{\cal H}={1\over 2}p_0^2+{1\over 2}q_0^2.
\label{e3.21}
\end{equation}
Next we use Eq.~(\ref{e3.15}) to rewrite Eq.~(\ref{e3.18}) in manifestly
Hermitian form:
\begin{eqnarray}
{d\over d\tau}{\cal B}=-{3\over 2}\left({\cal H}{\cal A}+{\cal A}{\cal H}\right)
\quad {\rm and} \quad {d\over d\tau}{\cal A}={3\over 2}\left({\cal H}{\cal B}
+{\cal B}{\cal H} \right).
\label{e3.22}
\end{eqnarray}

Suppose for a moment that we could replace ${\cal H}$ by the numerical constant
$C(0)$. We would then obtain the elementary classical system of coupled
differential equations in Eq.~(\ref{e2.20}). Because this system is {\em
linear} we could treat these operator differential equations classically and
ignore operator ordering. The solution to this system that satisfies the initial
conditions in Eq.~(\ref{e3.16}) would then be
\begin{eqnarray}
{\cal A}(\tau)=q_0\cos[3C(0)\tau]+p_0\sin[3C(0)\tau]~~{\rm and}~~
{\cal B}(\tau)=p_0\cos[3C(0)\tau]-q_0\sin[3C(0)\tau].
\label{e3.23}
\end{eqnarray}

This solution suggests the structure of the exact solution to the {\em operator}
differential equation system (\ref{e3.22}). The formal solution is a natural
quantum operator generalization of Eq.~(\ref{e3.23}) using Weyl-ordered products
of operators:
\begin{eqnarray}
{\cal A}(\tau) &=& {\cal W}[q_0\cos(3{\cal H}\tau)]+{\cal W}[p_0\sin(3{\cal H}
\tau)],\nonumber\\
{\cal B}(\tau) &=& {\cal W}[p_0\cos(3{\cal H}\tau)]-{\cal W}[q_0\sin(3{\cal H}
\tau)].
\label{e3.24}
\end{eqnarray}
The notation ${\cal W}[q_0 f({\cal H}\tau)]$ represents an operator ordering
defined as follows: First, expand the function $f({\cal H}\tau)$ as a Taylor
series in powers of the operator ${\cal H}\tau$. Then Weyl-order the Taylor
series term-by-term:
\begin{equation}
{\cal W}(q_0{\cal H}^n)\equiv{1\over 2^n}\left[(^n_0)q_0{\cal H}^n+(^n_1)
{\cal H}q_0{\cal H}^{n-1}+(^n_2){\cal H}^2q_0 {\cal H}^{n-2}+ ... +(^n_n)
{\cal H}^n q_0\right].
\label{e3.25}
\end{equation}
Using this definition it is straightforward to verify that Eq.~(\ref{e3.24}) is
indeed the {\em exact operator solution} to Eq.~(\ref{e3.22}) satisfying the
initial conditions in Eq.~(\ref{e3.16}).

Our objective is now to simplify the formal solution in Eq.~(\ref{e3.24}) and
to re-express it in closed form. Since sines and cosines are linear combinations
of exponential functions, we consider first the general problem of simplifying
the Weyl-ordered product
\begin{equation}
{\cal W}\left(q_0 e^{{\cal H}\tau}\right)={\cal W}\left[ q_0\left( 1+{{\cal H}
\tau}+{1\over 2!}({\cal H}\tau)^2+{1\over 3!}({\cal H}\tau)^3+...\right)\right].
\label{e3.26}
\end{equation}
For each power of $\tau$ we reorder the operators by commuting $q_0$
symmetrically to the left and to the right to maintain the Hermitian form
\begin{eqnarray}
{\cal W}(q_0)&=&q_0={1\over 2}(q_0 +q_0),\nonumber\\
{\cal W}(q_0{\cal H})&=&{1\over 2}(q_0{\cal H}+{\cal H}q_0)={\hbar\over 2}
\left(q_0{{\cal H}\over\hbar}+{{\cal H}\over\hbar}q_0\right),\nonumber\\
{\cal W}(q_0{\cal H}^2)&=&{1\over 4}\left( q_0{\cal H}^2+2{\cal H}q_0{\cal H}
+{\cal H}^2 q_0\right)={\hbar^2\over 2}\left[q_0\left({{\cal H}^2\over\hbar^2}-
{1\over 4}\right)+\left({{\cal H}^2\over\hbar^2}-{1\over 4}\right)q_0\right],
\nonumber\\
{\cal W}(q_0{\cal H}^3)&=&{1\over 8}\left( q_0{\cal H}^3+3{\cal H}q_0{\cal H}^2
+3{\cal H}^2 q_0{\cal H}+{\cal H}^3 q_0\right)\nonumber\\
&&\qquad\qquad={\hbar^3\over 2}\left[q_0\left({{\cal H}^3\over\hbar^3}
-{3\over 4}{{\cal H}\over\hbar}\right)+\left({{\cal H}^3\over\hbar^3}-{3\over 4}
{{\cal H}\over\hbar}\right)q_0\right],\nonumber\\
{\cal W}(q_0{\cal H}^4)&=&{1\over 16}\left( q_0{\cal H}^4+4{\cal H}q_0{\cal H}^3
+6{\cal H}^2q_0{\cal H}^2+4{\cal H}^3q_0{\cal H}+{\cal H}^4 q_0 \right)
\nonumber\\
&&\qquad\qquad ={\hbar^4\over 2}\left[q_0\left({{\cal H}^4\over\hbar^4}
-{3\over 2}{{\cal H}^2\over\hbar^2}+{5\over 16}\right)+\left({{\cal H}^4\over
\hbar^4}-{3\over 2}{{\cal H}^2\over\hbar^2}+{5\over 16}\right) q_0\right],
\label{e3.27}
\end{eqnarray}
and so on. This process defines a set of polynomials in the variable
${\cal H}/\hbar$
\cite{POLY}:
\begin{eqnarray}
&&1, \nonumber\\
&&{{\cal H}\over\hbar}, \nonumber\\
&&{{\cal H}^2\over\hbar^2}-{1\over 4}, \nonumber\\
&&{{\cal H}^3\over\hbar^3}-{3\over 4} {{\cal H}\over\hbar}, \nonumber\\ 
&&{{\cal H}^4\over\hbar^4}-{3\over 2}{{\cal H}^2\over\hbar^2}+{5\over 16},
\nonumber\\
&&{{\cal H}^5\over\hbar^5}-{5\over 2}{{\cal H}^3\over\hbar^3}+{25\over 16}
{{\cal H}\over\hbar},\nonumber\\
&&{{\cal H}^6\over\hbar^6}-{15\over 4}{{\cal H}^4\over\hbar^4}+{75\over 16}
{{\cal H}^2\over\hbar^2}-{61\over 64}.
\label{e3.28}
\end{eqnarray} 
We identify these polynomials as Euler polynomials \cite{AS} in which the
argument is shifted by $1\over 2$: $E_n\left({{\cal H}\over\hbar}+{1\over 2}
\right)$. The generating function for these nonorthogonal polynomials is given
by
\begin{equation}
{2e^{\left({{\cal H}\over\hbar}+{1\over 2}\right)\tau}\over e^\tau +1}=
\sum_{n=0}^{\infty} {\tau^n \over n!}E_n\left({{\cal H}\over\hbar}
+{1\over 2}\right)\quad (|\tau|<\pi).
\label{e3.29}
\end{equation}
This generating function allows us to express the following Weyl-ordered product
compactly:
\begin{equation}
{\cal W}\left(q_0 e^{{\cal H}\tau}\right)=
{q_0 e^{{\cal H}\tau}+e^{{\cal H}\tau}q_0\over 2\cosh (\tau\hbar/2)}.
\label{e3.30}
\end{equation}

Using Eq.~(\ref{e3.30}) the cosines and sines of our quantum solution in
Eq.~(\ref{e3.24}) can also be written in compact form when we take combinations
of complex exponentials:
\begin{eqnarray}
{\cal W}[q_0\cos(3{\cal H}\tau)] &=& {q_0\cos (3{\cal H}\tau)+\cos(3{\cal H}
\tau) q_0 \over 2\cos (3\tau\hbar/2)},\nonumber\\
{\cal W}[q_0\sin(3{\cal H}\tau)] &=&
{q_0\sin (3{\cal H}\tau)+\sin(3{\cal H}\tau)q_0\over 2\cos (3\tau\hbar/2)}.
\label{e3.31}
\end{eqnarray}

Last, we substitute this result into the zeroth-order solution in
Eq.~(\ref{e3.13}) and obtain
\begin{eqnarray}
Q_0(t,\tau) &=& {q_0\cos(t+3{\cal H}\tau)+\cos(t+3{\cal H}\tau)q_0\over 2\cos
(3\tau\hbar/2)}+{p_0\sin(t+3{\cal H}\tau)+\sin(t+3{\cal H}\tau)p_0\over 2\cos
(3\tau\hbar/2)}\nonumber\\
&=& {q_0\cos(t+3{\cal H}\epsilon t)+\cos(t+3{\cal H}\epsilon t)q_0\over 2\cos
(3\epsilon t\hbar/2)}+{p_0\sin(t+3{\cal H}\epsilon t)+\sin(t+3{\cal H}\epsilon
t)p_0\over 2\cos (3\epsilon t\hbar/2)},
\label{e3.32}
\end{eqnarray}
where we have replaced $\tau$ by $\epsilon t$. The result in Eq.~(\ref{e3.32})
is the objective of our multiple-scale analysis. It is the quantum operator
analog of Eq.~(\ref{e2.21}). Indeed, in the limit as $\hbar\to 0$, we recover
the classical multiple-scale approximation in Eq.~(\ref{e2.21}). [To obtain this
result we impose the classical initial conditions $p_0=0$ and $q_0=1$, which
from Eq.~(\ref{e3.2}) give ${\cal H}={1\over 2}$.]

We interpret this multiple-scale approximation to the operator $q$ as follows:
In the classical case we identify the coefficient of the time variable $t$ as
a first-order approximation to the frequency shift. However, here the
coefficient of $t$ in Eq.~(\ref{e3.32}) is an {\em operator}. Thus, we have
derived an operator form of mass renormalization.

To understand this operator mass renormalization we must take the expectation
value of Eq.~(\ref{e3.32}) between states. Then by examining the time dependence
of this matrix element we can read off the energy-level differences of the
quantum system. It is easy to construct a set of states because the operators
$q_0$ and $p_0$ satisfy the commutation relation (\ref{e3.7}). Hence,
appropriate linear combinations of $q_0$ and $p_0$  may be used as raising and
lowering operators to generate a Fock space consisting of the states
$| n\rangle$. By construction, these states are eigenstates of the operator
${\cal H}$:
\begin{eqnarray}
{\cal H} | n\rangle = \left( n+{1\over 2}\right)\hbar | n\rangle.
\label{e3.33}
\end{eqnarray}

Let us take the expectation value of Eq.~(\ref{e3.32}) between the states 
$\langle n-1|$ and $| n\rangle$. Allowing the operator ${\cal H}$ to act to the
left and the right using Eq.~(\ref{e3.33}), we obtain
\begin{eqnarray}
\langle n-1| Q_0 | n\rangle &=& \langle n-1| q_0 | n\rangle
 {\cos\left[ t+3\left(n+{1\over 2}\right)\hbar\epsilon t\right]
+\cos\left[ t+3\left(n-{1\over 2}\right)\hbar\epsilon t\right]
\over 2\cos (3\epsilon t\hbar/2)}\nonumber\\
&&\qquad +\langle n-1| p_0 | n\rangle
 {\sin\left[ t+3\left(n+{1\over 2}\right)\hbar\epsilon t\right]
+\sin\left[ t+3\left(n-{1\over 2}\right)\hbar\epsilon t\right]
\over 2\cos (3\epsilon t\hbar/2)}\nonumber\\
&=& \langle n-1| q_0 | n\rangle \cos [t(1+3n\hbar\epsilon )]
 +\langle n-1| p_0 | n\rangle \sin [t(1+3n\hbar\epsilon )],
\label{e3.34}
\end{eqnarray}
and we can see that the energy-level differences of the quantum oscillator are
$1+3n\hbar\epsilon$.

Let us verify this result. If we set $\epsilon=0$ in Eq.~(\ref{e3.1}), we obtain
the harmonic oscillator, whose coordinate-space eigenfunctions $\psi_n(x)$ and
corresponding energy eigenvalues $E_n$ are
\begin{eqnarray}
\psi_n(x)=e^{-{1\over 4}x^2}{\rm He}_n(x)\quad{\rm and}\quad E_n=n+{1\over 2},
\label{e3.35}
\end{eqnarray}
where ${\rm He}_n(x)$ represents the Hermite polynomial. The first-order
perturbative correction to the energy eigenvalues is obtained by computing the
expectation value of the perturbation term in the Hamiltonian $H$:
\begin{eqnarray}
E_n=n+{1\over 2}+{\epsilon\hbar\over 4}{\int_{-\infty}^{\infty}dx\, e^{-{1
\over 2}x^2}[{\rm He}_n(x)]^2x^4\over\int_{-\infty}^{\infty}dx\, e^{-{1\over 2}
x^2}[{\rm He}_n(x)]^2}=n+{1\over 2}+{3\over 4}\epsilon\hbar (2n^2+2n+1)+{\rm O}(
\epsilon^2).
\label{e3.36}
\end{eqnarray}
If we now calculate the energy difference $E_n-E_{n-1}$ from Eq.~(\ref{e3.36}),
we obtain
\begin{eqnarray}
E_n-E_{n-1} = 1+3n\hbar\epsilon + {\rm O}(\epsilon^2),
\label{e3.37}
\end{eqnarray}
which verifies the result in Eq.~(\ref{e3.34}).

\section{RESUMMATION OF THE CONVENTIONAL PERTURBATION SERIES FOR THE QUANTUM
ANHARMONIC OSCILLATOR}
\label{s4}

Now we turn our attention from the Heisenberg equations of motion to the
Schr\"odinger equation and examine the behavior of the wave function. The
ground-state wave function $\psi(x)$ for the quantum anharmonic oscillator
satisfies the Schr\"odinger equation
\begin{eqnarray}
\left( -{d^2 \over dx^2} + {1\over 4}x^2 + {1\over 4}\epsilon x^4
-E(\epsilon) \right)\psi(x) = 0
\label{e4.1}
\end{eqnarray}
and obeys the boundary conditions
\begin{eqnarray}
\psi(\pm\infty) = 0.
\label{e4.2}
\end{eqnarray}

We can use WKB analysis to find the large-$x$ asymptotic behavior of the wave
function $\psi(x)$. A geometrical-optics approximation gives the controlling
factor (the exponential component of the leading asymptotic behavior) as
\begin{eqnarray}
e^{-\sqrt{\epsilon} \vert x \vert^3 /6}.
\label{e4.3}
\end{eqnarray}
This result is nonperturbative in the sense that WKB is valid independent of the
size of the parameter $\epsilon$. The question addressed in this section is
whether it is possible to reproduce this result using perturbation theory.

The conventional approach to solving Eq.~(\ref{e4.1}) using
Rayleigh-Schr\"odinger perturbation theory \cite{BW1,BW2} represents both the
eigenfunction and eigenvalue as power series in $\epsilon$:
\begin{eqnarray}
\psi(x)\sim\sum_{n=0}^\infty\epsilon^n y_n(x)\quad {\rm and}
\quad E(\epsilon)\sim\sum_{n=0}^\infty\epsilon^n E_n.
\label{e4.4}
\end{eqnarray}
Note that we use the asymptotic symbol $\sim$ because, as is well known, the
conventional perturbation series for the anharmonic oscillator diverges.

Equations (\ref{e4.1}) and (\ref{e4.2}) are homogeneous, so we are free to adopt
the normalization condition $\psi(0)=1$, which translates into
\begin{eqnarray}
y_0(0)=1\quad {\rm and}\quad y_n(0)=0\quad (n>0).
\label{e4.5}
\end{eqnarray}

The unperturbed (harmonic-oscillator) solutions corresponding to $\epsilon=0$
are
\begin{eqnarray}
y_0(x)=e^{- x^2 /4}\quad {\rm and}\quad E_0={1\over 2}.
\label{e4.6}
\end{eqnarray}

In Ref.~\cite{BW1} it is shown that for all $n$, $y_n(x)$ is a product of the
zeroth-order approximation to the wave function $y_0(x)$ given in
Eq.~(\ref{e4.6}) multiplied by a polynomial $P_n(x)$:
\begin{eqnarray}
y_n =e^{- x^2 /4}P_n(x).
\label{e4.7}
\end{eqnarray}
Substituting Eq.~(\ref{e4.7}) into Eq.~(\ref{e4.1}), we obtain the recursion
formula for the polynomials:
\begin{eqnarray}
P_n^{\prime \prime}(x)-x P_n^{\prime}(x) = {1\over 4}x^4 P_{n-1}(x) -
\sum_{j=0}^{n-1} P_j(x) E_{n-j}.
\label{e4.8}
\end{eqnarray}
The form of this recursion relation is generic. It is a typical recursive
structure that arises in all perturbative calculations; to wit, the homogeneous
part of this recursion relation is the same for all $n$ while the inhomogeneous
part contains all previous polynomials. It is this kind of recursive structure
that is responsible for successive orders of perturbation theory being
resonantly coupled. Here, the resonant coupling causes the degree of the
polynomials to grow with $n$; $P_n(x)$ is a polynomial of degree $2n$ in the
variable $x^2$:
\begin{eqnarray}
P_0(x) &=& 1,\nonumber\\
P_1(x) &=& -{3\over 2}\left({x\over 2}\right)^2-\left({x\over 2}\right)^4,
\nonumber\\
P_2(x) &=& {21\over 4}\left({x\over 2}\right)^2+{31\over 8}\left({x\over 2}
\right)^4+{13\over 6}\left({x\over 2}\right)^6+{1\over 2}\left({x\over 2}\right)
^8,\nonumber\\
P_3(x) &=& -{333\over 8}\left({x\over 2}\right)^2-{243\over 8}\left({x\over 2}
\right)^4-{271\over 16}\left({x\over 2}\right)^6-{47\over 8}\left({x\over 2}
\right)^8-{17\over 12}\left({x\over 2}\right)^{10}-{1\over 6}\left({x\over 2}
\right)^{12},\nonumber\\
P_4(x) &=& {30885\over 64}\left({x\over 2}\right)^2+{2777\over 8}\left({x
\over 2}\right)^4+{18461\over 96}\left({x\over 2}\right)^6+{9195\over 128}\left(
{x\over 2}\right)^8+{979\over 48}\left({x\over 2}\right)^{10}+{599\over 144}
\left({x\over 2}\right)^{12}\nonumber \\
&&+ {7\over 12}\left({x\over 2}\right)^{14}+{1\over 24}\left({x\over 2}\right)
^{16},\nonumber\\
P_5(x) &=& -{916731\over 128}\left({x\over 2}\right)^2-{651363\over 128}\left(
{x\over 2}\right)^4-{89673\over 32}\left({x\over 2}\right)^6-{69107\over 64}
\left({x \over 2}\right)^8-{250183\over 768}\left({x\over 2}\right)^{10}
\nonumber\\
&&- {29177\over 384}\left({x\over 2}\right)^{12}-{1325\over 96}\left({x\over 2}
\right)^{14}-{269\over 144}\left({x \over 2}\right)^{16}-{25\over 144}\left({x
\over 2}\right)^{18}-{1\over 120}\left({x \over 2}\right)^{20},\nonumber\\
P_6(x) &=& {65518401\over 512}\left({x\over 2}\right)^2+{23046319\over 256}
\left({x\over 2}\right)^4+{75770813\over 1536}\left({x\over 2}\right)^6+{2476011
\over 128}\left({x\over 2}\right)^8\nonumber\\
&&+ {9259481\over 1536}\left({x\over 2}\right)^{10}+{13796435\over 9216}\left(
{x\over 2}\right)^{12}+{77173\over 256}\left({x\over 2}\right)^{14}+{112483
\over 2304}\left({x\over 2}\right)^{16}\nonumber\\
&&+ {4055\over 648}\left({x\over 2}\right)^{18}+{349\over 576}\left({x\over 2}
\right)^{20}+{29\over 720}\left({x\over 2}\right)^{22}+{1\over 720}\left({x
\over 2}\right)^{24}.
\label{e4.9}
\end{eqnarray}
For $n>0$ the general form of the polynomial is
\begin{eqnarray}
P_n(x)=\sum_{k=1}^{2 n}C_{n,k}(-{1\over 2}x^2)^k.
\label{e4.10}
\end{eqnarray}
Note that by virtue of Eq.~(\ref{e4.5}) the polynomials $P_n(x)$ have {\em no}
constant term when $n>0$.

We can derive a formula for $E_n$ and a recursion relation for the coefficients
$C_{n,k}$ by substituting Eq.~(\ref{e4.10}) into Eq.~(\ref{e4.8}) to obtain for
$n>0$
\begin{eqnarray}
C_{n,1}&=&E_n,\nonumber \\
2kC_{n,k}+C_{n-1,k-2}&=&-(k+1)(2k+1)C_{n,k+1}+\sum_{j=1}^{n-1}C_{j,k}C_{n-j,1}.
\label{e4.11}
\end{eqnarray}

The coefficients $C_{n,k}$ form a triangular array in the sense that the degree
of the polynomials increases with $n$. The convolution in Eq.~(\ref{e4.11})
makes this recursion relation highly nontrivial. However, for all $n$ it is
possible to find $C_{n,2n}$, the coefficient of the {\em highest power} of $x$.
This is an exact analog of finding the coefficient of the highest power of $t$
(most secular term) in $n$th order in perturbation theory for the classical
anharmonic oscillator. By setting $k=2n$ in Eq.~(\ref{e4.11}), we see that
$C_{n,2n}$ satisfies the simple linear recursion relation
\begin{eqnarray}
4nC_{n,2n} + C_{n-1,2n-2} =0,
\label{e4.12}
\end{eqnarray}
whose solution is
\begin{eqnarray}
C_{n,2n} =  \left ( -{1\over 4} \right )^n {1\over n!}.
\label{e4.13}
\end{eqnarray}

To study the behavior of the wave function $\psi(x)$ for large $x$, we
approximate $\psi(x)$ by resumming the perturbation series in Eq.~(\ref{e4.4})
and keeping just the {\em highest power} of $x$ in every order. Using
Eq.~(\ref{e4.13}) we obtain a simple exponential approximation to $\psi(x)$:
\begin{eqnarray}
\sum_{n=0}^\infty \epsilon^n e^{-x^2 / 4} { (-1)^n \over n! 16^n}
x^{4 n} = e^{-x^2/4} e^{- \epsilon x^4 /16}.
\label{e4.14}
\end{eqnarray}
 For the classical anharmonic oscillator this approach also gives an exponential
approximation [see Eq.~(\ref{e2.12})]. However, the classical and quantum
anharmonic oscillators are evidently quite different; although we have summed
the most secular terms to all orders in perturbation theory, the result in
Eq.~(\ref{e4.14}) is {\em not} the correct behavior of the wave function
$\psi(x)$ for large $x$ because it decays to zero too rapidly. The correct
behavior from WKB theory is an exponential of a cubic [see Eq.~(\ref{e4.3})] and
not an exponential of a quartic!

Can we improve our estimate of $\psi(x)$ by including $C_{n,2n-1}$, the
coefficient of the {\em next highest power} of $x$, in our summation? This
coefficient satisfies the inhomogeneous difference equation obtained by setting
$k=2n-1$ in Eq.~(\ref{e4.11}):
\begin{eqnarray}
(4n-2)C_{n,2n-1}+C_{n-1,2n-3}=-2n(4n-1)C_{n,2n}=
-2n(4n-1)\left(-{1\over 4}\right )^n {1\over n!}.
\label{e4.15}
\end{eqnarray}
The solution to this inhomogeneous equation is
\begin{eqnarray}
C_{n,2n-1} = -\left ( -{1\over 4} \right )^n {1\over n!}{n\over 3}(4n+5).
\label{e4.16}
\end{eqnarray}
If we include this formula in the resummation to all orders, we obtain the
result in Eq.~(\ref{e4.14}) now multiplied by a polynomial:
\begin{eqnarray}
e^{-x^2/4}e^{-\epsilon x^4 /16}\left( 1-{3\over 8}\epsilon x^2+{1\over 96}
\epsilon^2 x^6\right).
\label{e4.17}
\end{eqnarray}

We have again failed to obtain the correct large-$x$ behavior of $\psi(x)$.
Nevertheless, let us continue to reorganize the perturbation series. Setting
$k=2n-2$ in Eq.~(\ref{e4.11}), we get
\begin{eqnarray}
(4n-4)C_{n,2n-2}+C_{n-1,2n-4}=-(2n-1)(4n-3)C_{n,2n-1} + {3\over 4}C_{n-1,2n-2},
\label{e4.18}
\end{eqnarray}
whose solution is
\begin{eqnarray}
C_{n,2n-2}=\left( -{1\over 4}\right)^n{1\over n!}{n(n-1)\over 18}(16n^2+64n+87).
\label{e4.19}
\end{eqnarray}
The next few coefficients are:
\begin{eqnarray}
C_{n,2n-3}&=&-\left(-{1\over 4}\right)^n{1\over n!}{n(n-1)\over 162}(
64n^4+400n^3+764n^2-433n+390),\nonumber\\
C_{n,2n-4}&=&\left(-{1\over 4}\right)^n{1\over n!}{n(n-1)(n-2)\over 1944}
\nonumber \\
&&\qquad\times (256n^5+2816n^4+11744n^3+18304n^2+34209n+70029),\nonumber\\
C_{n,2n-5}&=&-\left(-{1\over 4}\right)^n{1\over n!}{n(n-1)(n-2)\over 29160}
\nonumber\\
&&\qquad \times
(1024n^7+14592n^6+74368n^5+106080n^4+45316n^3\nonumber\\
&&\qquad\qquad +143073n^2-2392803n+3967380),\nonumber\\
C_{n,2n-6}&=&\left(-{1\over 4}\right)^n{1\over n!}{n(n-1)(n-2)(n-3)\over 524880}
\nonumber\\
&&\qquad \times
(4096n^8+86016n^7+718080n^6+2799360n^5+6702384n^4\nonumber\\
&&\qquad\qquad +16486704n^3+16745975n^2 +180087585n+415966860).
\label{e4.20}
\end{eqnarray}
Evidently, the coefficient $C_{n,2n-j}$ has the general form of a polynomial in
$n$ of degree $2j$ multiplied by $(-4)^{-n}/n!$.

When we sum these coefficients to all orders, we obtain a totally new sequence
of polynomials (displayed below in square brackets) in the variable
$\epsilon x^4$:
\begin{eqnarray}
e^{- x^2 /4} e^{-\epsilon x^4 /16}  \bigg\{ 1 &+&\bigg[ -{3\over 2}\epsilon
\left( {x\over 2} \right)^2 + {2\over 3}\epsilon^2
\left( {x\over 2} \right)^6 \bigg] \nonumber\\
&+&\bigg[{31\over 8}\epsilon^2\left({x\over 2}\right)^4-2\epsilon^3\left({x\over
2}\right)^8+{2\over 9}\epsilon^4\left({x\over 2}\right)^{12}\bigg] \nonumber\\
&+&\bigg[ {21 \over 4}\epsilon^2 \left( {x\over 2} \right)^2
-{187\over 16}\epsilon^3 \left( {x\over 2} \right)^6
+{73\over 12}\epsilon^4 \left( {x\over 2} \right)^{10}
-\epsilon^5 \left( {x\over 2} \right)^{14}
+{4\over 81}\epsilon^6 \left( {x\over 2} \right)^{18}\bigg] \nonumber\\
&+&\bigg[ -{243\over 8} \epsilon^3 \left( {x\over 2} \right)^4
+{5307\over 128}\epsilon^4 \left( {x\over 2} \right)^8
-{58\over 3}\epsilon^5 \left( {x\over 2} \right)^{12}
+{133\over 36}\epsilon^6 \left( {x\over 2} \right)^{16}\nonumber\\
&&\qquad -{8\over 27}\epsilon^7 \left( {x\over 2} \right)^{20}
+{2\over 243}\epsilon^8 \left( {x\over 2} \right)^{24}\bigg] \nonumber\\
&+&\bigg[ -{333\over 8} \epsilon^3 \left( {x\over 2} \right)^2
+{14465\over 96}\epsilon^4 \left( {x\over 2} \right)^6
-{39493\over 256}\epsilon^5 \left( {x\over 2} \right)^{10}
+{12463\over 192}\epsilon^6 \left( {x\over 2} \right)^{14}\nonumber\\
&&\qquad -{313\over 24}\epsilon^7 \left( {x\over 2} \right)^{18}
+{211\over 162}\epsilon^8 \left( {x\over 2} \right)^{22}
-{5\over 81}\epsilon^9 \left( {x\over 2} \right)^{26}
+{4\over 3645}\epsilon^{10} \left( {x\over 2} \right)^{30}\bigg] \nonumber\\
&+&\bigg[ {2777\over 8} \epsilon^4 \left( {x\over 2} \right)^4
-{46891\over 64}\epsilon^5 \left( {x\over 2} \right)^8
+{5444579\over 9216}\epsilon^6 \left( {x\over 2} \right)^{12}
-{14497\over 64}\epsilon^7 \left( {x\over 2} \right)^{16}\nonumber\\
&&\qquad +{79357\over 1728}\epsilon^8 \left( {x\over 2} \right)^{20}
-{833\over 162}\epsilon^9 \left( {x\over 2} \right)^{24}
+{307\over 972}\epsilon^{10} \left( {x\over 2} \right)^{28}\nonumber\\
&&\qquad -{4\over 405}\epsilon^{11} \left( {x\over 2} \right)^{32} +{4\over
32805}\epsilon^{12} \left( {x\over 2} \right)^{36}\bigg] + ... \bigg\}.
\label{e4.21}
\end{eqnarray}
The original form of the perturbation series in Eqs.~(\ref{e4.4}) and
(\ref{e4.7}) has undergone a remarkable transmutation. The original polynomials
$P_n(x)$ have been absorbed and the wave function $\psi(x)$ is now represented
as a more elaborate exponential multiplying a new class of polynomials. [We can,
of course, generate these polynomials directly from the Schr\"odinger equation
(\ref{e4.1}) using a recursion relation similar to that in Eq.~(\ref{e4.8}).]

Let us perform a further resummation procedure. That is, for this new set of
polynomials we determine the coefficient of the highest power of $x$, 
\begin{eqnarray}
\left({2\over 3}\right)^n \left({x\over 2}\right)^{6n}
{1\over n!}\epsilon^{2n},
\label{e4.22}
\end{eqnarray}
the coefficient of the second-highest power of $x$,
\begin{eqnarray}
-{9\over 4} \left({2\over 3}\right)^n \left({x\over 2}\right)^{6n-4}
{1\over n!}\epsilon^{2n-1} n^2,
\label{e4.23}
\end{eqnarray}
the coefficient of the third-highest power of $x$,
\begin{eqnarray}
{9\over 32} \left({2\over 3}\right)^n \left({x\over 2}\right)^{6n-8}
{1\over n!}\epsilon^{2n-2} n(n-1)(9n^2-3n+1),
\label{e4.24}
\end{eqnarray}
the coefficient of the fourth-highest power of $x$,
\begin{eqnarray}
-{27\over 128} \left({2\over 3}\right)^n \left({x\over 2}\right)^{6n-12}
{1\over n!}\epsilon^{2n-3} n(n-1)(n-2)(9n^3-9n^2+7n+4),
\label{e4.25}
\end{eqnarray}
and so on.

Having discovered these formulas we can now sum over $n$. This further
reorganization of the perturbation series gives a {\em new} approximation to
$\psi(x)$ as an exponential multiplied by yet another set of polynomials, this
time in the variable $\epsilon^2 x^6$:
\begin{eqnarray}
e^{-x^2/4}e^{-\epsilon x^4/16}e^{\epsilon^2 x^6/96}\bigg\{ 1 &-&\bigg[{3\over 2}
\epsilon \left( {x\over 2}\right)^2 +\epsilon^3 
\left( {x\over 2}\right)^8  \bigg] \nonumber\\
&+&\bigg[ {31\over 8}\epsilon^2 \left({x\over 2}\right)^4
+{7\over 2} \epsilon^4 \left({x\over 2}\right)^{10}
+{1\over 2}\epsilon^6 \left({x\over 2}\right)^{16} \bigg] \nonumber\\
&-&\bigg[ {187\over 16}\epsilon^3  \left({x\over 2}\right)^6
+{277\over 24} \epsilon^5 \left({x\over 2}\right)^{12}
+{11\over 4}\epsilon^7 \left({x\over 2}\right)^{18}
+{1\over 6}\epsilon^9 \left({x\over 2}\right)^{24} \bigg]\nonumber\\
&+& ... \bigg\}.
\label{e4.26}
\end{eqnarray}

For these new polynomials the term containing the highest power of $x$ has the
form
\begin{eqnarray}
(-1)^n \left({x\over 2}\right)^{8n} {1\over n!}\epsilon^{3n},
\label{e4.27}
\end{eqnarray}
and the term containing the second-highest power of $x$ has the form
\begin{eqnarray}
{1\over 2}(-1)^n\left({x\over 2}\right)^{8n-6}{1\over n!}\epsilon^{3n-2}n(4n-1).
\label{e4.28}
\end{eqnarray}

Again, we reorganize the perturbation series by summing over all $n$. This
resummation gives yet another representation for $\psi(x)$ as an exponential
multiplied by a new set of polynomials, this time in the variable $\epsilon^3
x^8$:
\begin{eqnarray}
e^{-x^2/4}e^{-\epsilon x^4/16}e^{\epsilon^2 x^6/96} e^{-\epsilon^3 x^8/256}
\bigg\{&1&
+\bigg[-{3\over 2}\epsilon \left( {x\over 2}\right)^2 +2 \epsilon^4
\left( {x\over 2}\right)^{10}  \bigg] + ... \bigg\}.
\label{e4.29}
\end{eqnarray}

For these polynomials we identify the term containing the highest power of $x$,
\begin{eqnarray}
2^n\left({x\over 2}\right)^{10n} {1\over n!}\epsilon^{4n},
\label{e4.30}
\end{eqnarray}
and we reorganize the perturbation series again by summing over all $n$. This
resummation gives a new exponential and a new set of polynomials, this time in
the variable $\epsilon^4 x^{10}$:
\begin{eqnarray}
e^{-x^2/4}e^{-\epsilon x^4/16}e^{\epsilon^2 x^6/96} e^{-\epsilon^3 x^8/256}
e^{\epsilon^4 x^{10}/512} \bigg\{1 + ... \bigg\}.
\label{e4.31}
\end{eqnarray}

We can continue this process indefinitely. With each iterative reorganization of
the perturbation series we generate a new exponential multiplied by new
polynomials. However, we have not attained our original goal of deriving the
exponential behavior in Eq.~(\ref{e4.3}) from the weak-coupling perturbation
series in powers of $\epsilon$. Indeed, it seems impossible for our approach to
succeed because at each stage in the reorganization of the perturbation series
the variable $x$ appears only in {\em even} powers. How can we ever obtain the
exponential of a $\em cubic$?

There is a simple and direct answer to this question. We merely recognize that
the exponent in Eq.~(\ref{e4.31}) is the beginning of a binomial series:
\begin{eqnarray}
-{x^2\over 4}-{1\over 16}\epsilon x^4+{1\over 96}\epsilon^2 x^6-{1\over 256}
\epsilon^3 x^8 + {1\over 512} \epsilon^4 x^{10} + ... = {1\over 6\epsilon}
\left[ 1-(1+\epsilon x^2)^{3/2}\right].
\label{e4.32}
\end{eqnarray}
If we now let $x$ be large $(\epsilon x^2>>1)$, we recover the asymptotic
behavior in Eq.~(\ref{e4.3}):
\begin{eqnarray}
{\rm exp}\left\{ {1\over 6\epsilon} \left[ 1-(1+\epsilon x^2)^{3/2}\right]
\right\} \sim e^{-\sqrt{\epsilon} \vert x \vert^3 /6}\quad (|x|\to\infty).
\label{e4.33}
\end{eqnarray}

Let us examine more deeply what happens at each reorganization of the
perturbation series. At the first resummation we sum over terms of the form
$\epsilon^n x^{4n}$ and treat $\epsilon x^4$ as small. Thus, while $x$ may be
large compared with 1, it cannot be too large; $x$ must satisfy the asymptotic
bound $x<<\epsilon^{-1/4}$. At the next reorganization we sum over terms of the
form $\epsilon^{2n}x^{6n}$ and treat $\epsilon^2 x^6$ as small. This resummation
is valid in the larger region $x<<\epsilon^{-1/3}$. At the next level we sum
over terms of the form $\epsilon^{3n}x^{8n}$ and treat $\epsilon^3 x^8$ as
small. Hence, this resummation is valid in a still larger region, $x<<\epsilon^
{-3/8}$. At the $j$th iteration the range of $x$ increases to $x<<\epsilon^{j
\over 2j+2}$. Clearly, as $j\to\infty$, we obtain an estimate of the wave
function $\psi(x)$ that is valid for $x$ as large as $\epsilon^{-1/2}$. It is
only when $x$ is this large that the $\epsilon x^4$ term in the Schr\"odinger
equation (\ref{e4.1}) becomes comparable in size to the $x^2$ term! Thus, after
a finite number of reorderings of the perturbation series, we cannot expect to
reproduce exactly the exponential behavior in Eq.~(\ref{e4.3}).

Nevertheless, at each stage of the resummation process we observe precursors of
the exponential behavior in Eq.~(\ref{e4.3}). To demonstrate this, we examine
the structure of the exponent in Eq.~(\ref{e4.31}), in which we factor out
the first term:
\begin{eqnarray}
-{x^2\over 4} \left( 1 + {1\over 4}\epsilon x^2 - {1\over 24} \epsilon^2 x^4 +
{1\over 64}\epsilon^3 x^6 -{1\over 128}\epsilon^4 x^8 + ...\right).
\label{e4.34}
\end{eqnarray}
At the first iteration, which is valid for $x<<\epsilon^{-1/4}$, we neglect all
terms in Eq.~(\ref{e4.34}) beyond ${1\over 4}\epsilon x^2$ (because they are
small compared with 1) and replace this series by
\begin{eqnarray}
-{x^2\over 4} \left( 1+{1\over 2} \epsilon x^2 \right)^{1/2}.
\label{e4.35}
\end{eqnarray}
At the second iteration, which is valid for $x<<\epsilon^{-1/3}$, we neglect all
terms in Eq.~(\ref{e4.34}) beyond $-{1\over 24}\epsilon^2 x^4$ and replace
the series by
\begin{eqnarray}
-{x^2\over 4} \left( 1+ \epsilon x^2+{5\over 24}\epsilon^2 x^4 \right)^{1/4}.
\label{e4.36}
\end{eqnarray}
At the next two iterations we have
\begin{eqnarray}
-{x^2\over 4} \left( 1+ {3\over 2}\epsilon x^2 + {11\over 16}\epsilon^2 x^4
+{3\over 32}\epsilon^3 x^6 \right)^{1/6}
\label{e4.37}
\end{eqnarray}
and
\begin{eqnarray}
-{x^2\over 4} \left( 1+ 2\epsilon x^2 + {17\over 12}\epsilon^2 x^4
+{5\over 12}\epsilon^3 x^6+{47\over 1152}\epsilon^4 x^8 \right)^{1/8}.
\label{e4.38}
\end{eqnarray}

Now, let $x$ be large. In each of the above formulas we obtain a cubic in $x$!
Moreover, this cubic is multiplied by $-\sqrt{\epsilon}$ and the numerical
coefficients of the cubic approach ${1\over 6}$:
\begin{eqnarray}
{1\over 4 \sqrt{2}}     &=& {1\over 5.65685},\nonumber\\
{1\over 4 (24/5)^{1/4}} &=& {1\over 5.92066},\nonumber\\
{1\over 4 (32/3)^{1/6}} &=& {1\over 5.93469},\nonumber\\
{1\over 4 (1152/47)^{1/8}} &=& {1\over 5.96663}.
\label{e4.39}
\end{eqnarray}
It is now clear how it is possible to obtain a cubic in $x$ from only even
powers of $x$.

Having given a detailed discussion of the exponential prefactor that emerges at
each stage of the resummation process, we conclude this section by describing 
the structures of the polynomials that arise at every stage. Let $N$ represent
the number of resummations that we have performed. Then for each value of $N$
we can express the wave function $\psi(x)$ as an exponential multiplied by a
sum over polynomials. We have established that after $N$ resummations the 
exponential factor has the form
\begin{eqnarray}
{\rm exp}\left[-{x^2\over 4}\sum_{n=0}^N {\Gamma\left({3\over 2}\right)\over
\Gamma\left({3\over 2}-n\right)(n+1)!}(\epsilon x^2)^n\right].
\label{e4.40}
\end{eqnarray}
At each stage $N$ in the resummation process the argument of the polynomials
changes so we denote the argument by $z_N$, where
\begin{eqnarray}
z_N\equiv\epsilon^N\left({x^2\over 4}\right)^{N+1}.
\label{e4.41}
\end{eqnarray}
In general, the wave function is a product of the exponential factor multiplied
by a power series in the variable
\begin{eqnarray}
y\equiv{1\over 4}\epsilon x^2.
\label{e4.42}
\end{eqnarray}
We represent the coefficient of $y^n$ at the $N$th resummation by the notation
${\cal P}_n^{(N)} (z_N)$. Thus, 
\begin{eqnarray}
\psi(x)={\rm exp}\left[-{x^2\over 4}\sum_{n=0}^N{\Gamma\left({3\over 2}\right)
\over\Gamma\left({3\over 2}-n\right)(n+1)!} (\epsilon x^2)^n\right]
\sum_{n=0}^{\infty}y^n{\cal P}_n^{(N)}(z_N).
\label{e4.43}
\end{eqnarray}

We now describe the structure of ${\cal P}_n^{(N)}(z_N)$. For sufficiently large
$N$, $N>n-2$, ${\cal P}_n^{(N)}$ is a polynomial of degree $n$. However, when
$N\leq n-2$, negative powers are present. The largest negative power of $z_N$ is
$z_N^{-(n-N-1)}$. Thus, in the range from $N=0$ to $N=n-2$, ${\cal P}_n^{(N)}
(z_N)$ contains all powers of $z_N$ from $z_N^{-(n-N-1)}$ to $z_N^n$. But when
$N>n-2$, ${\cal P}_n^{(N)}(z_N)$ contains all powers of $z_N$ from $z_N^0$ to
$z_N^n$.

We now give some formulas for the functions ${\cal P}_n^{(N)}$. For all $N$,
${\cal P}_0^{(N)}$ is extremely simple:
\begin{eqnarray}
{\cal P}_0^{(N)}(z_N)=1.
\label{e4.44}
\end{eqnarray}
We have found a closed-form general expression for ${\cal P}_1^{(N)}$ valid for
all $N$:
\begin{eqnarray}
{\cal P}_1^{(N)}(z_N)=-{3\over 2}-2(-1)^N {(2N)!\over N!\,(N+2)!}z_N.
\label{e4.45}
\end{eqnarray}
The closed-form expression for ${\cal P}_2^{(N)}$ is more complicated:
\begin{eqnarray}
{\cal P}_2^{(N)}(z_N)=\left\{ \begin{array}{cc}
{21\over 4}z_0^{-1}+ {31\over 8} +{13\over 6}z_0+{1\over 2}z_0^2 & (N=0),\\
{31\over 8}-(-1)^N {(11N+13)(2N)!\over N!(N+3)!}z_N
+2\left[{(2N)!\over N!(N+2)!}\right]^2 z_N^2 & (N>0).\end{array}\right.
\label{e4.46}
\end{eqnarray}
The expression for ${\cal P}_3^{(N)}$ is still more complicated:
\begin{eqnarray}
{\cal P}_3^{(N)}(z_N)=\left\{ \begin{array}{cc}
 -{333\over 8}z_0^{-2}  - {243\over 8}z_0^{-1} 
- {271 \over 16}  - {47 \over 8} z_0 - {17 \over 12} z_0^2 
- {1 \over 6}z_0^3 & (N=0),\\
{21\over 4}z_1^{-1}-{187\over 16}+{73\over 12}z_1-z_1^2+{4\over 81}z_1^3 & (N=1)
,\\
-{187\over 16}+b_N z_N+c_N z_N^2
-(-1)^N{4\over 3}\left[{(2N)!\over N!(N+2)!}\right]^3 z_N^3 & (N>1),
\end{array}\right.
\label{e4.47}
\end{eqnarray}
where $b_N$ and $c_N$ are rational numbers.

Ginsburg and Montroll \cite{GM} attempted to approximate the wave function for
the anharmonic oscillator by a sequence of exponentials of fractional powers of
polynomials. In this respect their work is superficially similar to ours.
However, they did not generate these polynomials from the Rayleigh-Schr\"odinger
perturbation series; rather, they attempted to {\em fit} the small-$\epsilon$
and large-$x$ behaviors of $\psi(x)$ and then to deduce formulas for the
eigenvalues. Their work represented the wave function $\psi(x)$ by an
exponential function only, whereas in our work there are always polynomials
${\cal P}$ multiplying the exponential structure. These polynomials will never
disappear, even after an infinite number of resummations, because they express
the {\em physical-optics} correction to the geometrical-optics approximation to
the wave function. The exponential in Eq.~(\ref{e4.3}) is only the
geometrical-optics approximation.

\section{MULTIPLE-SCALE PERTURBATION THEORY APPLIED TO THE SCHR\"ODINGER
EQUATION FOR THE ANHARMONIC OSCILLATOR}
\label{s5}

In this section we attempt to obtain the large-$x$ asymptotic behavior of the
wave function $\psi(x)$ by applying MSPT directly to the Schr\"odinger equation
(\ref{e4.1}). We will see that multiple-scale analysis gives the results of the
previous section but it avoids the need to resum the perturbation series.

Recall that in Sec.~\ref{s4} we showed that the asymptotic form of the wave
function as the exponential of a cubic could only be obtained after an infinite
number of resummations. Since multiple-scale analysis is equivalent to resumming
the most secular terms in the perturbation series, we expect that at every order
of MSPT we will find a result similar in structure to the forms of the previous
section. Thus, we expect that leading-order MSPT based on just two scales will
fail to generate the correct cubic asymptotic behavior of the wave-function in
Eq.~(\ref{e4.3}). However, we expect that as we include more and more scales in
the MSPT we will obtain a series like that in Eqs.~(\ref{e4.32}) and
(\ref{e4.33}).

We will see that the application of MSPT to the Scr\"odinger equation
(\ref{e2.1}) is not straightforward. MSPT is conventionally applied to systems
that reduce to classical harmonic oscillators when the perturbation parameter
vanishes so that long scales are equal to the short scale multiplied by
increasing powers of the perturbation parameter $\epsilon$. The multiple-scale
analysis that we perform here is unusual because, as we have already see in
Sec.~\ref{s4}, the long scales are now proportional to increasing powers of
$x^2$ as well as $\epsilon$. In the following generalization of MSPT we will
show how the definition of the long scales relative to the short scale $x$ can
be deduced from the formalism and that secularity has a natural analogue in the
behavior of the wave function for large $x$.

To illustrate our procedure we begin by performing a multiple-scale analysis
involving only two scales. (A higher-order analysis will be given later on.) We
assume a perturbative solution to Eq.~(\ref{e4.1}) of the form
\begin{eqnarray}
\psi(x)=G_0(x,\xi_1)+\epsilon G_1(x,\xi_1)+{\rm O}(\epsilon^2),
\label{e5.1}
\end{eqnarray}
where the short scale is $x$ and the long scale $\xi_1$ is an {\em unknown}
function of $x$: $\xi_1=\epsilon f_1(x)$. (We use the subscript $1$ to indicate
that this is the {\em first} of a hierarchy of longer and longer scales.) This
assumption is reminiscent of the method of stretched coordinates \cite{MS}.

Using the chain rule we calculate the first two derivatives of $\psi(x)$:
\begin{eqnarray}
{d\psi\over dx} &=& {\partial G_0\over\partial x}+\epsilon\left({\partial G_0
\over\partial\xi_1}{df_1\over dx}+{\partial G_1\over\partial x}\right)+{\rm O}
(\epsilon^2),\nonumber\\
{d^2\psi\over dx^2} &=& {\partial^2 G_0\over\partial x^2}+\epsilon\left( 
{\partial^2 G_0\over\partial\xi_1\partial x}{df_1\over dx}+{\partial G_0\over
\partial\xi_1}{d^2f_1\over dx^2}+{\partial^2G_1\over\partial x^2}\right)+{\rm O}
(\epsilon^2).
\label{e5.2}
\end{eqnarray}

Next, we substitute Eqs.~(\ref{e5.1}) and (\ref{e5.2}) into the Schr\"odinger
equation and collect coefficients of powers of $\epsilon$. To order $\epsilon^0$
we obtain the Schr\"odinger equation for the quantum harmonic oscillator:
\begin{eqnarray}
-{\partial^2 G_0\over\partial x^2}+\left({1\over 4}x^2-{1\over 2}\right)G_0=0.
\label{e5.3}
\end{eqnarray}
The solution to this equation that is normalizable is
\begin{eqnarray}
G_0(x,\xi_1)=A(\xi_1)e^{-{1\over 4}x^2},
\label{e5.4}
\end{eqnarray}
where $A(\xi_1)$ is an unknown function of the long scale $\xi_1$.

To order $\epsilon^1$ we obtain
\begin{eqnarray}
-{\partial^2 G_1\over\partial x^2}+\left({1\over 4}x^2-{1\over 2}\right)G_1=2{
\partial^2 G_0\over\partial\xi_1\partial x}{df_1\over dx}+{\partial G_0\over
\partial\xi_1}{d^2f_1\over dx^2}-{1\over 4}x^4G_0+E_1G_0,
\label{e5.5}
\end{eqnarray}
where $E_1$ is the first-order correction to the energy eigenvalue.

Substituting the solution for $G_0$ into Eq.~(\ref{e5.5}) gives
\begin{eqnarray}
-{\partial^2 G_1\over\partial x^2}+\left({1\over 4}x^2-{1\over 2}\right)G_1=
e^{-{1\over 4}x^2}\left[-{dA\over d\xi_1}\left(x{df_1\over dx}-{d^2f_1\over
dx^2}\right)-A\left({1\over 4}x^4-E_1\right)\right].
\label{e5.6}
\end{eqnarray}

We now make the crucial argument in this procedure. By analogy with conventional
MSPT we do not want $G_1$ to satisfy an inhomogeneous equation driven by its
homogeneous solution $G_0$ because this will give a contribution to $G_1$ that
may decay less quickly than $G_0$ for large $x$. For example, if the square
bracket in Eq.~(\ref{e5.6}) were a constant, then $G_1$ would grow like
$G_0\log x$ for large $x$. We thus require that the square bracket on the right
side of Eq.~(\ref{e5.6}) vanish identically. Of course, one cannot give a
logical argument that this expression must vanish. Rather, we exploit the
functional degree of freedom that we introduced in Eq.~(\ref{e5.1}) when we
replaced a function of one variable by a function of two variables and simply
{\em demand that the square bracket vanish identically}.

We proceed as one does when performing a separation of variables for a partial
differential equation. The vanishing of the expression in square brackets gives
{\em two} ordinary differential equations, one for $f_1$ and one for $A$. Apart
from constants of integration that are irrelevant because they scale out, we
obtain
\begin{eqnarray}
f_1(x)={1\over 16}x^4+{3\over 8}x^2\qquad{\rm and}\qquad E_1={3\over 4},
\label{e5.7}
\end{eqnarray}
which fully defines the scale $\xi_1=\epsilon\left({1\over 16}x^4+{3\over 8}x^2
\right)$. Note that $E_1$ computed here is the usual perturbative correction to
the eigenvalue determined from Eq.~(\ref{e4.8}). Also, we have
\begin{eqnarray}
A(\xi_1)=e^{-\xi_1}.
\label{e5.8}
\end{eqnarray}
Finally, we use the expression for $\xi_1$ in terms of $x$ to rewrite the
solution for $G_0$:
\begin{eqnarray}
G_0=e^{-{1\over 4}x^2\left(1+{3\over 2}\epsilon\right)-{1\over 16}\epsilon x^4}.
\label{e5.9}
\end{eqnarray}

Observe that this generalized multiple-scale procedure leads to a solution that
incorporates not only the next term in the expansion of the exponent [see
Eq.~(\ref{e4.32})] but also contains a correction to the ${1\over 4}x^2$ term.
If we expand the exponent, we see that this correction gives the {\em constant
term} in the polynomial ${\cal P}_1^{(N)}(z_N)$ in Eq.~(\ref{e4.45}).

We now perform the multiple-scale approximation to higher order. We will see
that our multiple-scale procedure determines the new long-scale variables
required to describe the behavior of the wave function for large $x$. Moreover,
the higher-order calculation predicts exactly the higher-order corrections to
the energy eigenvalue. We generalize the first-order calculation by introducing
more long-scale variables $\xi_2=\epsilon^2f_2(x)$, $\xi_3=\epsilon^3f_3(x)$,
and so on. The functions $f_n$ are determined from the requirement that the
right side of the differential equation for $G_n$ vanish. This requirement
yields the new dependence of $G_0$ on $\xi_n$.

To order $\epsilon^2$ this procedure yields
\begin{eqnarray}
\xi_2=\epsilon^2\left({1\over 96}x^6+{11\over 64}x^4+{21\over 8}x^2\right)
\qquad{\rm and}\qquad E_2=-{21\over 8},
\label{e5.10}
\end{eqnarray}
and
\begin{eqnarray}
G_0(x)=e^{-{1\over 4}x^2\left(1+{3\over 2}\epsilon-{21\over 4}\epsilon^2\right)
-{1\over 16}\epsilon x^4\left(1-{11\over 4}\epsilon\right)+{1\over 96}\epsilon^2
x^6}.
\label{e5.11}
\end{eqnarray}
To order $\epsilon^3$ we have
\begin{eqnarray}
\xi_3=\epsilon^3\left({1\over 256}x^8+{21\over 192}x^6+{45\over 32}x^4+{333\over
32}x^2\right)\qquad{\rm and}\qquad E_3={333\over 16},
\label{e5.12}
\end{eqnarray}
and
\begin{eqnarray}
G_0(x)=e^{-{1\over 4}x^2\left(1+{3\over 2}\epsilon-{21\over 4}\epsilon^2+{333
\over 8}\epsilon^3\right)-{1\over 16}\epsilon x^4\left(1-{11\over 4}\epsilon+
{45\over 2}\epsilon^2\right)+{1\over 96}\epsilon^2 x^6\left(1-{21\over 2}
\epsilon\right)-{1\over 256}\epsilon^3 x^8}.
\end{eqnarray}
Observe the pattern that develops. Every new order in MSPT reproduces a new term
in the expansion of the exponent in Eq.~(\ref{e4.32}) together with
$x$-dependent corrections to the wave function in the exponential. To recover
the sequence polynomials ${\cal P}_1$, ${\cal P}_2$, ${\cal P}_3$, $\ldots$,
which are discussed in Sec.~\ref{s4} one need only expand the exponentials.

\section*{ACKNOWLEDGEMENTS}
\label{s6}
We thank T.~T.~Wu for a very helpful discussion. One of us, CMB, thanks the
Fulbright Foundation, the PPARC, and the U.S.~Department of Energy for financial
support. One of us, LMAB, thanks JNICT-{\it Programa Praxis XXI} for financial
support under contract No.~BD 2243/92. This work was supported in part by the
European Commission under the Human Capital and Mobility Programme, contract
No.~CHRX-CT94-0423.


\begin{references}
\bibitem{MASS} We have set $m^2=1$ in $H(p,q)$. Thus, $p$ and $q$ have the same
dimension.
\end{references}
\end{document}